\documentclass[tightenlines,draft,showpacs,eqsecnum,superscriptaddress,floats,prd,aps,twocolumn]{revtex4}
\usepackage{amsmath,amssymb,amsfonts}
\usepackage{graphicx}

\begin{document}
\preprint{UTBRG-2001-03, gr-qc/0203XXX}

\title{Reconstruction of Black Hole Metric Perturbations from Weyl Curvature}

\author{Carlos O. Lousto}
\affiliation{Department of Physics and Astronomy, 
The University of Texas at Brownsville,
Brownsville, Texas 78520}
\affiliation{Instituto de Astronom\'{\i}a y F\'{\i}sica del Espacio--CONICET,
Buenos Aires, Argentina}
\author{Bernard F. Whiting}
\affiliation{Department of Physics, PO Box 118440, Universitity of Florida,
Gainesville, FL 32611-8440}

\date{\today}

\begin{abstract} 
Perturbation theory of rotating black holes is usually described in
terms of Weyl scalars $\psi_4$ and $\psi_0$, which each satisfy
Teukolsky's complex master wave equation and respectively represent
outgoing and ingoing radiation. On the other hand metric perturbations
of a Kerr hole can be described in terms of (Hertz-like) potentials
$\Psi$ in outgoing or ingoing {\it radiation gauges}. In this paper we
relate these potentials to what one actually computes in perturbation
theory, i.e $\psi_4$ and $\psi_0$.  We explicitly construct these
relations in the nonrotating limit, preparatory to devising a
corresponding approach for building up the perturbed spacetime of a
rotating black hole.  We discuss the application of our procedure to
second order perturbation theory and to the study of radiation
reaction effects for a particle orbiting a massive black hole.
\end{abstract}

\pacs{04.25.Nx, 04.30.Db, 04.70.Bw}
\maketitle

\section{Introduction}

The spherically symmetry of a Schwarzschild black hole background allows for a
multipole decomposition of metric perturbations, even in the time domain. 
These were studied originally by Regge and Wheeler~\cite{Regge57} for
odd-parity  perturbations and by Zerilli~\cite{Zerilli70a} for the even-parity
case.  Moncrief~\cite{Moncrief74} has given a gauge-invariant formulation of
the problem, in terms of the three-metric perturbations. The two degrees of
freedom of the gravitational field are described in terms of two waveforms,
$\phi^\pm$ satisfying a simple wave equation
\begin{equation}\label{rtrw}
-\frac{\partial^2\phi^\pm_{(\ell m)}}{\partial t^2}
+\frac{\partial^2\phi^\pm_{(\ell m)}}{\partial r*^2}-
V_{\ell^\pm}(r)\phi^\pm_{(\ell m)}=0 \; .
\end{equation}
Here $r^*\equiv r+2M\ln(r/2M-1)$, and $V_{\ell^\pm}(r)$ are the 
Zerilli and Regge--Wheeler potentials respectively.

Given the solution to the wave equation (\ref{rtrw}) one can reconstruct
explicitly both the even and odd parity
metric perturbations of a Schwarzschild background
in the Regge--Wheeler gauge~\cite{Lousto99b,Nicasio:2000ge}.
This permits a complete description
of the perturbative spacetime. But the Regge--Wheeler gauge is 
unfortunately not asymptotically
flat, and in order to extract radiation information for a second order
perturbative expansion one has to perform a new gauge
transformation~\cite{Gleiser:1998rw}.  Moreover, the desirable 
properties of the the Regge--Wheeler gauge being unique
and invertible are lost in the case the background is a
{\it rotating} black hole, i.e. represented by the Kerr metric,
where no effective multipole decomposition is yet known.

There is an independent formulation of the perturbation problem
derived from the Newman-Penrose formalism~\cite{Newman62a} that is
valid for perturbations of rotating black holes.\cite{Teukolsky73}
This formulation fully exploits the null structure of black holes to
decouple the curvature perturbation equations into a single wave
equation that, in Boyer-Lindquist coordinates $(t,r,\theta,\varphi)$,
can be written as:
\begin{eqnarray}
&&\ \ \Biggr\{\left[ a^2\sin^2\theta -\frac{(r^2+a^2)^2}\Delta \right]
\partial_{tt}-\frac{4Mar}\Delta \partial_{t\varphi } \nonumber \\
&&-2s\left[ (r+ia\cos
\theta )-\frac{M(r^2-a^2)}\Delta \right] \partial_t  \nonumber \\
&&\ \ +\,\Delta^{-s}\partial_r\left( \Delta^{s+1}\partial_r\right) +
\frac 1{\sin \theta }\partial_\theta \left( \sin \theta \partial_\theta
\right) \nonumber \\
&& +\left( \frac 1{\sin^2\theta }-\frac{a^2}\Delta \right) 
\partial_{\varphi \varphi }
+\,2s\left[ \frac{a(r-M)}\Delta +\frac{i\cos \theta }{\sin^2\theta }
\right] \partial_\varphi  \nonumber \\
&&-\left( s^2\cot^2\theta -s\right) \Biggr\}\psi
=4\pi \Sigma T \;,  \label{master}
\end{eqnarray}
where $M$ is the mass of the black hole, $a$ its angular momentum per unit
mass, $\Sigma  \equiv r^2+a^2\cos^2\theta$, and $\Delta 
\equiv  r^2-2Mr+a^2$.
The source term $T$ is built up from the energy-momentum
tensor~\cite{Teukolsky73}.  Gravitational perturbations, 
corresponding to $s=\pm2$, are compactly 
described in terms of contractions of the Weyl tensor with a null 
tetrad.  The components of the tetrad (also given in Ref.\ \cite{Teukolsky73}) 
are appropriately chosen along the repeated principal null directions 
of the background spacetime [see Eq.\ (\ref{tetrad}) below].  The resulting
gauge and (infinitesimally) tetrad invariant components of the Weyl curvature are
given by
\begin{equation}\label{psi}
\psi=\left\{ 
\begin{array}{ll}
\rho^{-4}\psi_4\equiv -\rho^{-4}C_{n\bar mn\bar m} & {\rm for}~~s=-2
\\ 
\psi_0\equiv -C_{lmlm} & {\rm for}~~s=+2~
\end{array}
\right. , 
\end{equation}
where an overbar means complex conjugation and $\rho$ is given in Eq.\
(\ref{spincoeff}) below. 
Asymptotically, the leading behavior of the field $\psi$
represents either the outgoing radiative part of the perturbed Weyl tensor, 
($s=-2$), or the ingoing radiative part, ($s=+2$).

The components of the Boyer--Lindquist null tetrad for the Kerr
background are given by
\begin{subequations}\label{tetrad}
\begin{eqnarray}
   (l^{\alpha}) &=& \left( \frac{r^2+a^2}{\Delta},1,0,\frac{a}{\Delta} 
                  \right), \\
   (n^{\alpha}) &=& \frac{1}{2\;\!(r^2+a^2\cos^2\theta)} \,
                  \left( r^2+a^2,-\Delta,0,a \right), \\
   (m^{\alpha}) &=& \frac{1}{\sqrt{2}(r+ia\cos\theta)}\,
                  \left( ia\sin\theta,0,1,i/\sin\theta \right).\quad\quad 
\end{eqnarray}
\end{subequations}
With this choice of the tetrad the non-vanishing spin coefficients are
\begin{eqnarray}\label{spincoeff}
\rho&=&-\frac{1}{(r-ia\cos\theta)},\quad
\beta=-\bar{\rho}\frac{\cot\theta}{2\sqrt{2}},\nonumber\\
\pi&=&ia\rho^2\frac{\sin\theta}{\sqrt{2}},\quad
\tau=-ia\rho\bar{\rho}\frac{\sin\theta}{\sqrt{2}},\nonumber\\
\mu&=&\rho^2\bar{\rho}\ \frac{\Delta}{2},\quad
\alpha=\pi-\bar{\beta},\nonumber\\
\gamma&=&\mu+\rho\bar{\rho}\frac{(r-M)}{2},
\end{eqnarray}
and the only non-vanishing Weyl scalar in the background is
\begin{equation}\label{psi2} 
\psi_2=M\rho^3.
\end{equation}

Analogously to the Zerilli-Regge-Wheeler waveforms, $\psi_4$ can be
directly used to compute energy and momentum radiated at infinity, but
it remains to relate it to metric perturbations.
Chandrasekhar~\cite{Chandrasekhar83} studied a way to to obtain metric
perturbations of the Kerr metric, but it was proved by Price and
Ipser~\cite{Price91} that this choice is not a proper gauge, namely is
an incomplete constraint on the coordinates.

Chrzanowski~\cite{Chrzanowski:1975wv}
generalized work of Cohen and Kegeles~\cite{Cohen:1974} on
Hertz potentials to the gravitational perturbations of the Kerr metric.
In Ref.\ \cite{Chrzanowski:1975wv}
explicit expressions are given for {\it homogeneous}
metric perturbations in the frequency domain. 
Wald~\cite{Wald:1978vm} subsequently showed that the expressions given in
Ref.\ \cite{Chrzanowski:1975wv} do
not lead to {\it real} metric perturbations.
Cohen and Kegeles\ \cite{Kegeles:1979an}
then corrected their expressions and gave explicit equations
(see Sec.\ \ref{formulation})
relating metric perturbations to a gravitational Hertz potential,
$\Psi$ that fulfills
Eq.\ (\ref{master}), but that is different from $\psi_4$ or $\psi_0$.
In those works no explicit method was given for determining $\Psi$ in
any specific astrophysical problem.

In Sec.\ \ref{solution} we provide the explicit formulae 
relating a gravitational Hertz potential to $\psi_4$ or $\psi_0$
in the time domain,
hence to the given initial data defining the astrophysical model one
wants to evolve (See Ref.\ \cite{Campanelli98c} for the 3+1 decomposition of
the Weyl scalars).
These allow one to define the {\it outgoing} and {\it ingoing} radiation gauges
that are asymptotically flat at future infinity and regular on the event
horizon
respectively. Such gauges have been found\ \cite{Campanelli99} especially well suited for computing second order
perturbations of a Kerr hole and, once generalized
in presence of matter, can provide a first step
toward computing radiation reaction (self-force)
corrections\ \cite{Lousto99b} to the trajectory of a particle orbiting
a rotating black hole. 

Vacuum second order perturbations have a direct application to the
close limit expansion of the final merger stage of binary black holes
with comparable masses.  Perturbation theory in conjunction with
limited full numerical simulations has proved to be an extremely powerful
tool to compute waveforms from binary black holes from near the
innermost stable circular orbit~\cite{Baker:2001nu,Baker:2002qf}.  The
combination of radiation reaction correction plus second order
perturbations provides a formidable tool for computing gravitational
radiation from binary black hole--neutron star systems, and is
particularly relevant to the computation of template banks for ground
based interferometers such as LIGO/VIRGO/GEO about to enter on-line,
as well as space missions such as LISA, sensitive to supermassive
black hole binary systems. Thus in order to incorporate these
improvements to our theoretical predictions it is imperative to know
how to build up explicit metric perturbations around a Kerr
background.

\section{Formulation of the problem}\label{formulation}

We use the notation
\begin{equation}
g_{\mu\nu}=g_{\mu\nu}^{Kerr}+h_{\mu\nu}
\end{equation}
to describe metric perturbations.

\subsection{ingoing and outgoing radiation gauges}\label{IORG}

Chrzanowski \cite{Chrzanowski:1975wv},
and Cohen and Kegeles \cite{Kegeles:1979an}
found two convenient gauges that allow one to invert the
metric perturbations in terms of potentials $\Psi_{IRG}$ or $\Psi_{ORG}$
satisfying the same wave equations as the Weyl scalars $\rho^{-4}\psi_4$
or $\psi_0$ respectively.

In the {\it ingoing radiation} gauge (IRG) we have
\begin{eqnarray}\label{incalibre}
h_{ll}&=&h_{\mu\nu}l^\mu l^\nu=0;\quad h_{ln}=h_{\mu\nu}l^\mu n^\nu=0,\nonumber\\
h_{lm}&=&h_{\mu\nu}l^\mu m^\nu=0;\quad h_{l\overline{m}}=h_{\mu\nu}l^\mu
\overline{m}^\nu=0,\nonumber\\
h_{m\overline{m}}&=&h_{\mu\nu}m^\mu\overline{m}^\nu=0,
\end{eqnarray}
and the homogeneous (for vacuum) metric components can be written,
in the time domain, in terms of solutions to the wave equation for
$\rho^{-4}\psi_4$ only, as follows:\footnote{We have 
made the choice that $\epsilon=0$.} 
\begin{subequations}  \label{inmetrica}
\begin{eqnarray}
h_{nn}^{\ IRG}&=&-\left\{(\delta +\overline{\alpha }+3\beta-\tau)
(\delta +4\beta+3\tau)\right\}\nonumber\\
&&(\Psi_{IRG})+ c.c.\\
h_{\overline{m}\overline{m}}^{\ IRG}&=&
-\left\{(D-\rho)(D+3\rho)\right\}(\Psi_{IRG}),\\
h_{(n\overline{m})}^{\ IRG}&=&-\frac12\left\{(\delta-\overline{\alpha }+
3\beta-\overline{\pi }-\tau)(D+3\rho)+\right.\nonumber\\
&&\left.(D+\overline{\rho}-\rho)
(\delta +4\beta+3\tau)\right\}(\Psi_{IRG}),
\end{eqnarray}
\end{subequations}
where $c.c.$ stands for the complex conjugate part of the whole object
to ensure that the metric be real~\cite{Wald:1978vm,Kegeles:1979an}.
Note that in this gauge the metric potential has the properties of
being transverse $(h_{\mu\nu}l^\mu=0)$ and traceless $(h^{\mu}_\mu=0)$
at the future horizon and past null infinity. It is thus a suitable
gauge to study gravitational radiation effects near the event horizon.

The complementary (adjoint) gauge to the ingoing radiation gauge is the
{\it outgoing radiation} gauge (ORG), which can be obtained from the IRG
upon exchange of the tetrad
vectors $l\leftrightarrow n$, $\overline{m}\leftrightarrow m$ and
the appropriate renormalization. It satisfies:

\begin{equation}
h_{nn}=0=h_{ln}=0=h_{nm}=0=h_{n\overline{m}}=0=
h_{m\overline{m}} .  \label{outcalibre}
\end{equation}
The metric potential has now the property of being transverse 
$(h_{\mu\nu}n^\mu=0)$ and traceless $(h^{\mu}_\mu=0)$ at the
past horizon and future null infinity. It is thus an example of a suitable
asymptotically flat gauge in which to directly compute radiated energy and 
momenta at infinity.

In this gauge, the homogeneous metric components can be written in terms of
solutions to the wave equation for $\psi_0$, as
\begin{subequations} \label{outmetrica}
\begin{eqnarray}
h_{ll}^{\ ORG}&=&-\rho^{-4} \left\{(\overline
\delta -3\alpha -\overline\beta +5\pi) (\overline\delta -4\alpha+\pi)
\right\}\nonumber\\
&&(\Psi_{ORG})+ c.c.\\
h_{mm}^{\ ORG}&=&
-\rho^{-4} \left\{(\widehat\Delta+5\mu-3\gamma+\overline\gamma)
(\widehat\Delta+\mu-4\gamma)\right\}\nonumber\\
&&(\Psi_{ORG}),\\
h_{(lm)}^{\ ORG}&=&
-\frac12\rho^{-4} \Big\{(\overline\delta -3\alpha+\overline%
\beta+5\pi+\overline\tau) (\widehat\Delta+\mu-4\gamma)\nonumber\\
&&+(\widehat\Delta+5\mu-\overline\mu%
-3\gamma-\overline\gamma) (\overline\delta-4\alpha +\pi)
\Big\}\nonumber\\
&&(\Psi_{ORG}),
\end{eqnarray}
\end{subequations}
where the directional derivatives are
$D=l_{\ }^{\mu}\partial_\mu, \widehat\Delta=n_{\ }^{\mu}\partial_\mu,
\delta=m_{\ }^{\mu}\partial_\mu$, and the rest of the Greek 
letters represent the usual notation for spin coefficients.\cite{Newman62a}


\subsection{4th order equations for the potential}

From the evolution of the Teukolsky equation we can obtain
$\psi_4$ and $\psi_0$, the two gauge and tetrad invariant
objects representing outgoing and ingoing radiation respectively.
To relate the unknown potential $\Psi$ to them we use their
definitions\ (\ref{psi}) to obtain
\begin{eqnarray}
\psi_0&=&DDDD\left[\bar{\Psi}_{IRG}\right],\quad {\rm and}\label{psiIRG1}\\
\rho^{-4}\psi_4&=&\frac14\left[\bar{\cal L}\bar{\cal L}\bar{\cal L}\bar{\cal L}
\bar{\Psi}_{IRG}-12\rho^{-3}\psi_2\partial_t\Psi_{IRG}\right],\label{psiIRG2}
\end{eqnarray}
where $\bar{\cal L}=\bar{\eth}+ia\sin\theta\partial_t$ and
$\bar{\eth}=
-[\partial_\theta+s\cot\theta-i\csc\theta\partial_\varphi]$,
while for the outgoing radiation gauge we have
\begin{eqnarray}
\rho^{-4}\psi_4&=&\Delta^2
\widehat\Delta\widehat\Delta\widehat\Delta\widehat\Delta
\left[\Delta^2\bar{\Psi}_{ORG}\right],\quad {\rm and}\label{psiORG1}\\
\psi_0&=&\frac{1}{4}\left[{\cal L}{\cal L}{\cal L}{\cal L}
\bar{\Psi}_{ORG}+12\rho^{-3}\psi_2\partial_t\Psi_{ORG}\right].\quad\quad
\label{psiORG2}
\end{eqnarray}
In order to obtain an expression for the potentials in terms of
the known quantities $\psi_4$ or $\psi_0$, one has to invert a fourth
order differential equation where $\psi_4$ or $\psi_0$ act as source
terms. This will be the central task of our paper.



\section{Explicit solution for nonrotating black holes}\label{solution}

\subsection{Master equation}

The key observation here is that for $a=0$
the differential operator ${\cal L}$ acting on the potentials
in Eqs.\ (\ref{psiIRG2}) and Eqs.\ (\ref{psiORG2}) contains only
angular derivatives. Since for the spherically symmetric
background we can decompose the
potentials into spherical harmonics of spin weight $s$ and, from
\cite{Goldberg:1967}, we have
\begin{subequations}\label{eth}
\begin{eqnarray}
{}_{s}\bar{Y}_{\ell m}&=&(-)^{m+s}{}_{-s}Y_{\ell,-m},\\
\bar\eth\bar\eth\bar\eth\bar\eth\left[\ {}_{-2}\bar{Y}_{\ell m}\right]
&=&(-)^{m-2}\frac{(\ell+2)!}{(\ell-2)!}\ {}_{-2}Y_{\ell,-m},\quad\\
\eth\eth\eth\eth\left[\ {}_{2}\bar{Y}_{\ell m}\right]
&=&(-)^{m+2}\frac{(\ell+2)!}{(\ell-2)!}\ {}_{2}Y_{\ell,-m},\quad
\end{eqnarray}
\end{subequations}
we obtain a first order relationship among $\psi_4$ or $\psi_0$ and
the IRG or ORG potentials decomposed into multipoles
\begin{eqnarray}\label{psilm}
\left[\ \rho^{-4}\psi_4\right]^{\pm}&=&
\pm\frac{(\ell+2)!}{4(\ell-2)!}{\Psi}_{IRG}^{\pm}
-3M\partial_t\Psi_{IRG}^{\pm},\label{psi4IRG}\\
\psi_0^{\pm}&=&
\pm\frac{(\ell+2)!}{4(\ell-2)!}{\Psi}_{ORG}^{\pm}
+3M\partial_t\Psi_{ORG}^{\pm},\quad\quad\label{psi0IRG}
\end{eqnarray}
where we have used the notation 
\begin{eqnarray}\label{psipm}
\psi^\pm=\frac12[\ {\psi}^{\ell,m}\pm
(-)^{m}\overline{{\psi}^{\ell,-m}}\ ]
\end{eqnarray}
for all fields decomposed into multipoles.

Since $\Psi_{IRG}$ and $\Psi_{ORG}$ satisfy the master equation (\ref{master})
for spin $s=\mp2$ respectively we can eliminate from this equation
all time derivatives by replacing Eqs.\ (\ref{psi4IRG}) or (\ref{psi0IRG}) 
and its time derivatives into the Teukolsky equation. This leads to
the following equation for the IRG / ORG potentials, both represented here
by $\Psi$
\begin{eqnarray}\label{masterAS}
&&\Delta^{-s}\partial_r[\Delta^{s+1}\partial_r\Psi^{\pm}]
-\frac{r^4}{\Delta}(\Omega_{AS})^2\Psi^{\pm}\nonumber\\
&&\pm2sr(\Omega_{AS})(\frac{Mr}{\Delta}-1)\Psi^{\pm}\nonumber\\
&&-(\ell-s)(\ell+s+1)\Psi^{\pm}=F^{\pm}
\end{eqnarray}
where
\begin{equation}  \label{Psi}
\Psi=\left\{ 
\begin{array}{ll}
\Psi_{IRG} & {\rm for}~~s=-2
\\ 
\Psi_{ORG} & {\rm for}~~s=+2~
\end{array}
\right. , 
\end{equation}

\begin{equation}
\Omega_{AS}=\frac{1}{12M}\frac{(\ell+2)!}{(\ell-2)!},
\end{equation}
and the source term is
\begin{eqnarray}\label{F}
F^{\pm}&&:=-\frac{r^4}{3M\Delta} (\partial_t\psi^{\pm})\mp
\frac{r^4}{3M\Delta}(\Omega_{AS})({\psi}^{\pm})\nonumber\\
&&+2s\frac{r}{3M}(Mr/\Delta-1)(\psi^{\pm}).
\end{eqnarray}

The equation\ (\ref{masterAS}) is our fundamental equation to solve for the
potential in terms of the known fields $\psi_4$ or $\psi_0$ that appear in the
source terms. The key observations here are to note that, separately for the
``plus'' and ``minus'' parts: i) the left hand side of this equation contains the
Teukolsky operator in the {\it frequency} domain for the algebraically special
frequencies, $\omega=\pm i \Omega_{AS}$, and ii) the source terms in
(\ref{F}) are the Cauchy data for the Teukolsky operator in terms of $\psi_4$
or $\psi_0$, precisely as they would appear in a Laplace transform approach to
Eq.\ (\ref{master}) [See Eqs.\ (A2-A3) of Ref.\ \cite{Campanelli98a}]. Notably,
we have arrived to this equation working in the time domain, {\it without} any
frequency decomposition.


\subsection{solution}

The single frequency appearing on the right hand side of Eq.\ (\ref{master})
is precisely an algebraic special frequency.
Algebraically special perturbations of black holes excite gravitational
waves which are either purely ingoing or outgoing. Hence only one of 
$\psi_4$ {\it or} $\psi_0$ is non-zero while the other vanishes.
Chandrasekhar \cite{Chandrasekhar:1984}
has obtained the explicit form of the algebraic
special perturbations of the Kerr black hole.
This is a remarkable fact, because we can then use these analytic solutions
to the Teukolsky equation for algebraic special perturbations to
construct explicit solutions to (\ref{masterAS}).

For Schwarzschild, 
Chandrasekhar\ \cite{Chandrasekhar:1984} gives two algebraic special solutions
\begin{eqnarray}
y_1(r)&=&\left[1+\frac{\lambda r}{M}+\frac{\lambda^2 r^2}{3M^2}
+\frac{\lambda^2(\lambda+1) r^3}{9M^3}\right]e^{-\Omega_{AS}r^*},\quad\quad
\label{y1}\\
y_2(r)&=&\left[1-\frac{(\lambda+1) r}{3M}\right]r^2e^{+\Omega_{AS}r^*},
\label{y2}
\end{eqnarray}
where $\lambda=(\ell-1)(\ell+2)/2$.  Note that $y_2$ gives rise to an 
algebraically special solution for $\psi_4$, ($s=-2$), at frequency
$\omega^-_{AS}=-i\Omega_{AS}$ and for $\Delta^2 \psi_0$, ($s=+2$), at frequency 
$\omega^+_{AS}=+i\Omega_{AS}$, while $y_1$ gives rise to an algebraically
special solution for $\psi_4$ at frequency $\omega^+_{AS}$ and for 
$\Delta^2 \psi_0$ at frequency $\omega^-_{AS}$.

These two solutions satisfy the ``plus'' and ``minus''
parts of Eq.\ (\ref{masterAS}) for $\Psi_{IRG}^\pm$ ($s=+2$) and vice-versa
for $\Psi_{ORG}^\mp$ ($s=-2$).
A second set of independent solutions can be found: 
\begin{eqnarray}
z_1(r)&=&y_1(r)\int_{2M}^{r}\frac{W(r')}{y_1(r')^2}\ dr',\label{z1}\\
z_2(r)&=&y_2(r)\int_{\infty}^{r}\frac{W(r')}{y_2(r')^2}\ dr',\label{z2}
\end{eqnarray}
where the Wronskian of the solutions is
\begin{eqnarray}\label{W}
W(r):&=&y_{1,2}(r)\partial_rz_{1,2}(r)-z_{1,2}(r)\partial_ry_{1,2}(r)
\nonumber\\
&=&\Delta(r)=r(r-2M).
\end{eqnarray}
{\bf Note:}  $z_1$ and $z_2$ are not algebraically special solutions, although
they are each a homogeneous solution to their respective Teukolsky equation at
an algebraically special frequency.

Making use of these solutions, the explicitly expression for the
potential can be written as follows:
\begin{eqnarray}
\Psi_{IRG}^{+}(r)&=&-y_1(r)\int_{2M}^{r}\frac{z_1(r')F^{+}(r')}{\Delta(r')^2}
\ dr'\nonumber\\
&-&z_1(r)\int_{r}^{\infty}\frac{y_1(r')F^{+}(r')}{\Delta(r')^2}\ dr',
\quad\quad\quad\label{RpsiORG}\\
\Psi_{IRG}^{-}(r)&=&z_2(r)\int_{2M}^{r}\frac{y_2(r')F^{-}(r')}{\Delta(r')^2}
\ dr'\nonumber\\
&+&y_2(r)\int_{r}^{\infty}\frac{z_2(r')F^{-}(r')}{\Delta(r')^2}\ dr',
\quad\quad\quad\label{IpsiORG}
\end{eqnarray}
valid for each hypersurface where $\psi_0$ or $\psi_4$, entering in
$F$ given by Eq.~(\ref{masterAS}), are evaluated.

The equations for the ORG are obtained by exchanging ``plus'' and ``minus'' 
parts above (cf. Eq.\ (\ref{Rechi}) and (\ref{Imchi})), 
and by adopting the corresponding dependence of $F$ on 
the components of $\psi$ [See Eq.\ (\ref{psi})].



\section{Applications}\label{applications}

\subsection{Metric perturbations (IRG)}

Using the Regge-Wheeler notation \cite{Regge57} for metric perturbations,
conditions (\ref{incalibre}) read
\begin{eqnarray}
&&h_0^\ell(r,t)^{\rm (even,odd)}=-(1-2M/r)\,h_1^\ell(r,t)^{\rm (even,odd)},
\nonumber\\
&&G^\ell(r,t)=\frac{2}{\ell(\ell+1)}K^\ell(r,t),\nonumber\\
&&H_0^\ell(r,t)=H_2^\ell(r,t)=-H_1^\ell(r,t).
\end{eqnarray}

We can now explicitly compute the metric perturbations (\ref{inmetrica}) or
(\ref{outmetrica}) in terms of the computed $\psi_0$ or $\psi_4$:\footnote{For
Newman-Penrose scalars, as in Eqs. (\ref{h0par}) -- (\ref{h2impar}), even and
odd really refer to contributions to the real and imaginary parts of the whole
scalar, not specifically to the individual coefficients in a mode sum decomposition.}
\begin{eqnarray}\label{h0par}
&&[h_{(n\overline{m})}]^\ell\nonumber\\
&&=\left\{h_0^\ell(r,t)^{\rm (even)}-i 
h_0^\ell(r,t)^{\rm (odd)}\right\}
\frac{\sqrt{\ell(\ell+1)}}{\sqrt{2}r}\ {}_{-1}Y_{\ell m}\nonumber\\
&&=\left\{\left(\frac{y_1'}{ry_1}+\frac{C_\ell}{12M}
\frac{1}{(1-2M/r)}-\frac{2}{r^2}\right)[\Psi_{IRG}^+]\right.\nonumber\\
&&\left.-\frac{\Delta}{ry_1}\int_r^\infty\ \frac{y_1}{\Delta^2}[F^+]\ dr'
-\frac{[\rho^{-4}\psi_4^+]}{3M(r-2M)}\right.\nonumber\\
&&\left.+\left(\frac{y_2'}{ry_2}-\frac{C_\ell}{12M}
\frac{1}{(1-2M/r)}-\frac{2}{r^2}\right)[\Psi_{IRG}^-]\right.\nonumber\\
&&\left.+\frac{\Delta}{ry_2}\int^r_{2M}\ \frac{y_2}{\Delta^2} [F^-]\
dr' -\frac{[\rho^{-4}\psi_4^{-}]}{3M(r-2M)}\right\}\nonumber\\
&&\times\sqrt{2(\ell+2)(\ell-1)}\ {}_{-1}Y_{\ell m},
\end{eqnarray}
where $C_\ell=(l+2)!/(l-2)!$, and
\begin{eqnarray}
[h_{nn}]^\ell&=&(1-2M/r)\ H_0^\ell(r,t)\ {}_{0}Y_{\ell m}\nonumber\\
&=&\frac{\rho^2}{2}[\Psi_{IRG}^+]
\sqrt{\frac{(\ell+2)!}{(\ell-2)!}}\ {}_{0}Y_{\ell m},
\end{eqnarray}
\begin{eqnarray}
[h_{\overline{m}\overline{m}}]^\ell&=&
\frac12\left\{G(r,t)^\ell+i\frac{h_2(r,t)^\ell}{r^2}\right\}
\sqrt{\frac{(\ell+2)!}{(\ell-2)!}}\ {}_{-2}Y_{\ell m}
\nonumber\\
&=&[h_{\overline{m}\overline{m}}]^{\rm (even)}+
[h_{\overline{m}\overline{m}}]^{\rm (odd)}
\end{eqnarray}
where
\begin{eqnarray}
&&[h_{\overline{m}\overline{m}}]^{\rm (even)}=\Bigg\{
1/6\,{\frac {[\Psi_{IRG}^+]\left (12\,{M}^{2}+{r}^{2}C_{{l}}
\right ){y_1'}}{y_{{1}}r\left (r-2\,M\right )M}}\nonumber\\
&&-1/6\,{\frac {
\left (\Delta\right )\,\left (12\,{M}^{2}+{r}^{2}C_{{l}}\right)}
{y_{{1}}r\left (r-2\,M\right )M}}\int_r^\infty\ \frac{y_1}{\Delta^2} [F^+]\ dr'\nonumber\\
&&-2/3\,{\frac {{r}^{2}
[{\rho}^{-4}\partial_t\psi_{{4}}^+]}{\left (r-2\,M\right )^{
2}M}}-2/3\,{\frac {r{\frac {\partial }{\partial r}}[{\rho}^{-4}\psi_{{4}
}^+]}{\left (r-2\,M\right )M}}\nonumber\\
&&-1/18\,{\frac {[{\rho}^{-4}\psi_{{4}}^+]
\left (-36\,r+84\,M+{r}^{2}C_{{l}}\right )}{\left (r-2\,M\right )^{2}M}}\nonumber\\
&&+\frac {1}{72}\,\left ({r}^{3}{C_{{l}}}^{2}+12\,
r\left (6\,{l}^{2}+6\,l+7\,C_{{l}}-12\right ){M}^{2}\right.\nonumber\\
&&\left.-144\,\left (l+2
\right )\left(l-1\right ){M}^{3}-36\,{r}^{2}C_{{l}}M\right )\nonumber\\
&&\times\frac {[\Psi_{{{
\it IRG}}}^+]}{\left(r-2\,M\right )^{2}{M}^{2}r}
\Bigg\}\ {}_{-2}Y_{\ell m},
\end{eqnarray}
and
\begin{eqnarray}\label{h2impar}
&&[h_{\overline{m}\overline{m}}]^{\rm (odd)}=\Bigg\{
1/6\,{\frac {[\Psi_{{{\it IRG}}}^-]\left (12\,{M}^{2}-{r}^{2}C_{{l}}
\right ){y_2'}}{y_{{2}}r\left (r-2\,M\right )M}}\nonumber\\
&&+1/6\,{\frac {
\left (\Delta\right )\,\left (12\,{M}^{2}-{r}^{2}C_{{l}}\right )}
{y_2r\left (r-2\,M\right )M}}\int^r_{2M}\ \frac{y_2}{\Delta^2} [F^-]\
dr'\nonumber\\ &&-2/3\,{\frac {{r}^{2}
[{\rho}^{-4}\partial_t\psi_{{4}}^-]}{\left (r-2\,M\right )^{
2}M}}-2/3\,{\frac {r{\frac {\partial }{\partial
r}}[{\rho}^{-4}\psi_{{4} }^-]}{\left (r-2\,M\right )M}}\nonumber\\
&&-1/18\,{\frac {[{\rho}^{-4}\psi_{{4}}^-]
\left (-36\,r+84\,M-{r}^{2}C_{{l}}\right )}
{\left (r-2\,M\right )^{2}M}}\nonumber\\ &&+\frac {1}{72}\,\left
({r}^{3}{C_{{l}}}^{2}+12\, r\left (6\,{l}^{2}+6\,l-7\,C_{{l}}-12\right
){M}^{2}\right.\nonumber\\ &&\left.-144\,\left (l+2
\right )\left(l-1\right ){M}^{3}+36\,{r}^{2}C_{{l}}M\right )\nonumber\\
&&\times\frac {[\Psi_{{IRG}}^-]}{\left(r-2\,M\right )^{2}{M}^{2}r}
\Bigg\}\ {}_{-2}Y_{\ell m}.
\end{eqnarray}
In writing these we have explicitly lowered the order of the derivatives of the
potential by making use of the following identities
\begin{eqnarray}\label{Psiprime}
&&\partial_r\left([\Psi^{+}]\right)=\nonumber\\
&&\left(\frac{\partial_ry_1}{y_1}\right)[\Psi^+]-
\frac{\Delta}{y_1}\int_r^\infty\ \left(\frac{y_1}{\Delta^2}\right) [F^+]\ dr',\quad\quad\quad\\
&&\nonumber\\
&&\partial_r\left([\Psi^-]\right)=\nonumber\\
&&\left(\frac{\partial_ry_2}{y_2}\right)[\Psi^-]+
\frac{\Delta}{y_2}\int_{2M}^r\ \left(\frac{y_2}{\Delta^2}\right) [F^-]\ dr',\quad\quad\quad
\end{eqnarray}
and directly from Eq.\ (\ref{psi4IRG})
\begin{eqnarray}\label{Psidot}
\partial_t\Psi_{IRG}^\pm=
\pm\frac{(\ell+2)!}{12M(\ell-2)!}\Psi_{IRG}^\pm
-\frac{1}{3M\rho^{4}}\psi_4^\pm.\quad\quad
\end{eqnarray}


\subsection{Metric perturbations (ORG)}

Using the Regge-Wheeler notation \cite{Regge57} for metric perturbations
conditions (\ref{outcalibre}) read
\begin{eqnarray}
&&h_0^\ell(r,t)^{\rm (even,odd)}=(1-2M/r)\,h_1^\ell(r,t)^{\rm (even,odd)},
\nonumber\\
&&G^\ell(r,t)=\frac{2}{\ell(\ell+1)}K^\ell(r,t),\nonumber\\
&&H_0^\ell(r,t)=H_2^\ell(r,t)=H_1^\ell(r,t).
\end{eqnarray}

The explicit metric perturbations can be found directly from the previous
subsection, Eqs.\ (\ref{h0par})-(\ref{h2impar}) upon exchanges of the
tetrad contractions
$l\to n$ and $m\to\bar{m}$, and the consequent change in normalizations, $s$,
and potentials, as described throughout the paper.


\section{Discussion}\label{discussion}

We have explicitly computed the metric perturbations of a nonrotating
black hole in terms of the Weyl scalars $\psi_4$ and $\psi_0$ which
can be computed directly by solving the Teukolsky equation for any
appropriate astrophysical scenario, given the corresponding initial
data. In doing so we had to invert Eqs.\ (\ref{psiIRG2}) or\
(\ref{psiORG2}). This was performed making explicit use of the
multipole decomposition of the potential, Weyl scalars and metric.
The extension of this procedure to the rotating background is not
straightforward, but we have learned some key features: The algebraic
special solutions of the Teukolsky equation will play a crucial role
in finding the solution for the Hertz potential in terms of $\psi_4$
or $\psi_0$.  One can see this as follows. In order to invert Eq.\
(\ref{psiIRG2}) for the potential we first seek out solutions of the
homogeneous equation. Hence $\psi_4$ should vanish, but for the
solution to be nontrivial, $\psi_0$, given by Eq.\ (\ref{psiIRG1}), must
not vanish \cite{Wald:1973}.  These two conditions are precisely the
conditions that define the algebraically special solutions for the
potential satisfying the vacuum Teukolsky equation
(\ref{master}). These two solutions (in the time domain) should allow
one to build up the Kernel that inverts the fourth order equation
(\ref{psiIRG2}).  
An identical argument applies for the ORG
potential when working with Eqs.\ (\ref{psiORG1}) and (\ref{psiORG2}).

One main application of this formalism is to go beyond first
order perturbation theory and compute second order perturbations
of rotating black holes. In Ref.\ \cite{Campanelli99} there was developed a
formalism for the second order Teukolsky equation that takes
the form of the first order wave operator acting on the
second order piece of the Weyl scalar $\psi_4$, and a {\it source}
term build up as a quadratic combination of first order perturbations.
It is precisely to compute this source term that one needs the
explicit form of the metric perturbations. In Ref.\ \cite{Campanelli99}
it was found that to describe the emitted gravitational radiation the
ORG gauge is specially well suited. Hence one has to solve the
the first order Teukolsky equation for $\psi_4^{(1)}$ and
$\psi_0^{(1)}$, then later to build up the source of the second
order piece of the Weyl scalar $\psi_4^{(2)}$ (See Eq.~(9)
in Ref.~\cite{Campanelli99}).

A second important application of the reconstruction of metric
perturbations around Kerr background is to compute the self force
of a particle orbiting a massive black hole 
\cite{Mino:1997nk,Quinn:1997am} and to
compute corrected trajectories~\cite{Lousto99b} depending on the
perturbed metric and connection coefficients along the particle
world line. This task is left for a future paper.
While we know the form of the Teukolsky equation in the presence of
of perturbative matter around a Kerr hole (see Eq.\ (\ref{master})),
we need to generalize the equation satisfied by the potential and
the relationship between this potential and the metric perturbations,
i.e. the generalization of Eqs.\ (\ref{inmetrica}) and (\ref{outmetrica}).
In particular we know that not all of conditions (\ref{incalibre}) or
(\ref{outcalibre}) can hold in the presence of matter since they are then
incompatible with the Einstein equations~\cite{Barack:2001ph}.


\begin{acknowledgments}
We wish to thank H. Beyer, M. Campanelli and S. Detweiler for helpful
discussions.  This research has been supported in part by NSF Grant No.
PHY-9800977 (B.F.W.) with the University of Florida. C.O.L developed part of
this work in the Albert Einstein Institut, Germany, which B.F.W. also thanks
for hospitality.
\end{acknowledgments}

\appendix
\section{Relation between IRG/ORG potentials}\label{details}

We give here a relation expressing the potential
$\Psi_{ORG}$ in terms of the result for $\Psi_{IRG}$.  This
was not given in
\cite{Chrzanowski:1975wv,Cohen:1974,Wald:1978vm,Kegeles:1979an}.
We begin by defining a
field $\chi$, with spin weight $-2$, through: 
\begin{equation}\label{chidef}
\Psi_{IRG}=\frac{1}{4}\left[
\bar{\cal L}\bar{\cal L}\bar{\cal L}\bar{\cal L}
\bar{\chi}+12\rho^{-3}\psi_2\partial_t\chi\right],
\end{equation}
\noindent analogous to Eq.\ (\ref{psiORG2}).
The solution for $\Psi_{ORG}$ is:
\begin{equation}\label{psiORGf}
\Psi_{ORG}=DDDD\left[\bar{\chi}\right],
\end{equation}
also the relation (\ref{psiIRG1}) between $\psi_0$ 
and $\Psi_{IRG}$.  A potential degeneracy 
for algebraically special 
modes was discussed briefly in~\cite{Whiting:1989vc}.  
The explicit solution for the modes $\chi^{\ell m}$ can be written, in terms of
$\chi^{\pm}$ given by Eq. (\ref{psipm}), as:
\begin{eqnarray}
\chi^{+}(r)&=&z_2(r)\int_{2M}^{r}\frac{y_2(r')V^{+}(r')}{\Delta(r')^2}
\ dr'\nonumber\\
&+&y_2(r)\int_{r}^{\infty}\frac{z_2(r')V^{+}(r')}{\Delta(r')^2}\ dr',
\quad\quad\label{Rechi}\\
\chi^{-}(r)&=&-y_1(r)\int_{2M}^{r}\frac{z_1(r')V^{-}(r')}{\Delta(r')^2}
\ dr'\nonumber\\
&-&z_1(r)\int_{r}^{\infty}\frac{y_1(r')V^{-}(r')}{\Delta(r')^2}\ dr',
\quad\quad\label{Imchi}
\end{eqnarray}

\noindent cf. Eqs.\ (\ref{RpsiORG}) and (\ref{IpsiORG}) above.  
Here $V^\pm$ are given by:
\begin{eqnarray}\label{V}
V^\pm&&:=\frac{r^4}{3M\Delta} (\partial_t\Psi_{IRG}^\pm)\mp
\frac{r^4}{3M\Delta}(\Omega_{AS})
(\Psi_{IRG}^\pm)\nonumber\\
&&+4\frac{r}{3M}(Mr/\Delta-1)(\Psi_{IRG}^\pm),
\end{eqnarray}
similar to Eq.\ (\ref{F}), but note the sign differences.
\bibliographystyle{bibtex/prsty}
\bibliography{bibtex/references}
\thebibliography{potente4}
\end{document}